\documentclass[traditabstract,letter]{aa} 
\usepackage{graphicx}
\usepackage{natbib}
\bibpunct{(}{)}{;}{a}{}{,} 
\begin{document}
\title{Ground-based secondary eclipse detection of the very-hot
Jupiter OGLE-TR-56b\thanks{Based on data collected with the
6.5 meter Magellan Telescopes located at Las Campanas Observatory and the FORS2
imager at the VLT-Antu telescope in Paranal Observatory, ESO, Chile (program 081.C-0266).}
}

\author{David K. Sing\inst{1} and Mercedes L\'{o}pez-Morales\inst{2}\thanks{Hubble Fellow}}

   \offprints{D. K. Sing}

\institute{Institut d'Astrophysique de Paris, CNRS/UPMC, 98bis boulevard Arago, 75014 Paris, France\\ \email{sing@iap.fr}
 \and
Carnegie Institution of Washington, Dept. of Terrestrial Magnetism, 5241 Broad Branch Road NW, Washington, DC 20015, USA\\ \email{mercedes@dtm.ciw.edu}}

   \date{Received 31 October 2008/ Accepted 21 November 2008}

  \abstract{We report on the detection of the secondary eclipse of
the very-hot Jupiter \object{OGLE-TR-56b} from combined $z'$-band time
series photometry obtained with the VLT and Magellan telescopes.  We
measure a flux decrement of 0.0363$\pm$0.0091\% from the combined
Magellan and VLT datasets, which indicates a blackbody brightness
temperature of $2718^{+127}_{-107}$ K, a very low albedo, and a small
incident radiation redistribution factor, indicating a lack of
strong winds in the planet's atmosphere. The measured secondary depth
is consistent with thermal emission, but our precision is not
sufficient to distinguish between a black-body emitting planet, or
emission as predicted by models with strong optical absorbers such as
TiO/VO. This is the first time that thermal emission from an extrasolar
planet is detected at optical wavelengths and with ground-based
telescopes. }

\titlerunning{Ground-Based Secondary Eclipse Detection of OGLE-TR-56}
\authorrunning{Sing \& L\'{o}pez-Morales}
   \maketitle

%

\section{Introduction}

Very hot Jupiters (VHJs) are giant planets orbiting very close to
their host star, typically with periods of less than 2--3 days. Their
proximity to the stars makes those planets potentially very
hot, and it is therefore expected that they will emit detectable
amounts of thermal radiation both at optical and near-infrared
wavelengths \citep{2007ApJ...667L.191L,2008ApJ...678.1419F}.

Since the first detection of thermal emission from hot-Jupiters
\citep{2005Natur.434..740D}, all detections have been made
from space using the Spitzer telescope at wavelengths longer than 3.6
$\mu$m, despite numerous ground-based attempts
(e.g. \citealt{2007PASP..119..616K,2007MNRAS.378..148D,2007MNRAS.375..307S,2005MNRAS.363..211S}).
Anti-transit exoplanet atmospheric detections from the ground so far have
been elusive, in contrast to the recent ground-based primary
transit result by \cite{2008ApJ...673L..87R}. This is
unfortunate, as ground-based anti-transit detections can greatly
facilitate hot Jupiter atmospheric studies, and extend this field to
the optical and near-infrared regions of the spectrum.

\cite{2007ApJ...667L.191L} predicted the detectability of
thermal emission of OGLE-TR-56b in $z'$-band (0.9 $\mu$m). The authors found
that if the planet had a bolometric albedo $A_{B}$ = 0 and a negligible
atmospheric energy redistribution factor, the decrement in flux during
secondary eclipse could
be as large as 0.05$\%$ of the total light of the system. Of the hot
Jupiter planets known at that time, OGLE-TR-56b was
the best suited candidate for detection.

Here we present the results of the follow-up observations to test that
prediction. \S2 describes our observations and the data reduction and
analysis are presented in \S3.  In \S4 we summarize and discuss our results.
\section{Observations}

OGLE-TR-56b is a faint target (V=16.56) located in a very
crowded field.  These factors, combined with the small expected signal
of the secondary eclipse, make a detection challenging. Stable photometric
conditions and large aperture telescopes, which minimize stellar
blends and also provide high resolution time sampling of the target's light
curves, are therefore necessary.

We first monitored three secondary eclipse events of OGLE-TR-56b on
June 25th, 30th and July 2nd, 2008 UT using the FORS2 instrument at
the VLT-Antu telescope located at Paranal
Observatory in Chile.  FORS2 has a mosaic of two 2k$\times$4k E2V CCDs
with 15$\times$15 $\mu$m pixels.  We used the direct imaging mode of
the instrument with a Gunn-z filter, a high resolution collimator and
2$\times$2 binning, to obtain images with 0.125''/pixel and a field of
view (FoV) of 4.2'$\times$4.2' around the target.

Only the night of July 2nd produced high enough quality data to detect 
the secondary transit of OGLE-TR-56b. On that
night we benefited from both stable and good conditions, with an
average seeing of 0.72'' varying by $\pm$0.2'' during the run.  The
observations were taken between the Julian dates of 2454649.67536 and
2454649.88428, yielding 354 images with exposure times between 10 and
30 seconds and 26 second readout times.  Exposure times were adjusted
during the night to keep the counts in the central pixel of the target
PSF around half-well depth. The pointing was stable, with
the target centered to within $\pm$0.5 pixel in the CCD's x-direction
and $\pm$0.25 in the y-direction throughout the night.

The observations on June 25th suffer from poor and highly variable seeing
conditions, with seeing $>$1.3'' during the night, and could not
be used for this analysis.  The night of July 30th had good seeing,
between 0.6'' and 0.22'', but the sky conditions changed both
rapidly and dramatically during the in-eclipse phases, while remaining
stable out-of-eclipse. This affects the performance of our
de-trending algorithms (see \S3), which use the systematics found in
the out-of-eclipse portions of the light curve to correct the data
in-eclipse. The in- and out-of eclipse systematics proved to be
very different and impossible to remove.

We monitored another secondary eclipse event of OGLE-TR-56b on 
August 3rd, 2008 UT with the new MagIC-E2V instrument on the
Magellan-Baade telescope located at Las Campanas Observatory in Chile.
The MagIC-E2V is a two-amplifier CCD with 1024$\times$1024,
13$\times$13 $\mu$m pixels, producing a FoV of 38''$\times$38'' and a
resolution of 0.037'' per pixel in the 1$\times$1 binning
configuration.  All the exposures were taken with a SDSS $z'$ filter.
The observations span the Julian dates of 2454682.48142 to
2454682.69383 containing 253 images with 60 second integration times
and 5 second readout times.  The sky conditions were stable and good,
with an average seeing of 0.9'' which only varied by $\pm$0.1'' during
the run.


\section{Reduction and analysis}

Several stars, including the target, appear well isolated both in the
VLT and Magellan frames. This renders our analysis by aperture
photometry precise enough for this study.

 After correcting for bias and flat-fielding all images
using standard IRAF routines, we started the data analysis by
performing DAOPHOT-type aperture photometry, recording the flux from
all bright non-saturated stars in the FoV for a large and well sampled
range of apertures.  In the Magellan data, the sky background was
estimated using annuli centered around the apertures, and with
a large and well sampled range of inner and outer ring radii.  In the
VLT data, we estimated the sky background using an iterative clipping
method in the central third of the image, which contains the target
and reference stars, in a manner similar to the calculation of sky background 
in DAOPHOT.

\begin{figure}
 {\centering
  \includegraphics[width=0.37\textwidth,angle=90]{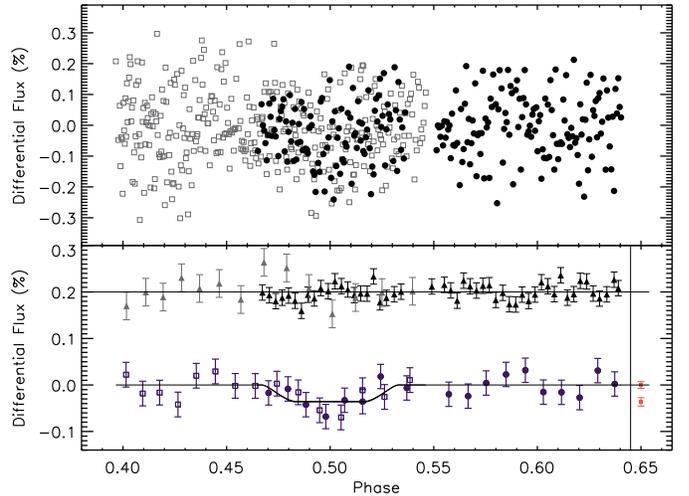}}
\caption[]{{\it Top}: OGLE-TR-56 de-trended differential light curves. Open 
squares and filled circles show, respectively, the VLT and Magellan 
data. {\it Bottom}:  VLT (squares) and Magellan (circles) light curves binned by a 
factor of 23 and 14 respectively. The blue light curve corresponds to OGLE-TR-56, and the black and grey
triangles to reference stars.  Models with and without the best fit 
secondary transit are shown as horizontal lines.  At phase 0.65 (red squares), we indicate the final values and uncertainties on the fit eclipse depth as well as the out-of-eclipse flux.}
\end{figure}

 The best comparison stars were selected separately in the Magellan
and VLT datasets by identifying the most stable (i.e. minimum standard
deviation) differential light curves between each star and
OGLE-TR-56. The individual preliminary light curves of both OGLE-TR-56
and the selected comparisons presented clear systematic trends, with
the VLT dataset more highly affected.  These trends can be attributed
to atmospheric effects, such as seeing, sky-background brightness and
airmass variations, or instrumental effects such as small changes of
the location of the stars in the detector. Those effects can be
modeled and reduced using de-trending algorithms commonly used in
transit light curve analyses.

A standard procedure in transit light curve analysis is to model
systematics using only portions of each star's light curve at
phases out-of-transit and then to apply the model to the in-transit data,
to prevent removing the transit signal itself. For data-sets with a
large number of light curves, systematic errors can be 
corrected by algorithms like SysRem
\citep{2005MNRAS.356.1466T}, which measures trends common to many stars
of the sample, removing each trend with a pass through the algorithm.

Our field is in a very crowded region toward the Galactic center,
limiting the number of useful unblended reference stars and the
performance of iterative algorithms.  Therefore, we de-trend the light
curves in two steps.  First we
modeled and removed a single initial linear trend common to the target and
reference stars using an implementation of the SysRem
algorithm \citep{2005MNRAS.356.1466T}, which seeks to minimize the
expression $\sum(r_{ij}-c_ia_j)^2/\sigma^2_{ij}$, where $r_{ij}$ is
the average-subtracted stellar magnitude for the $i$th star of the
$j$th image, $\sigma$ is the uncertainty of $r_{ij}$, $c_i$ is an
epoch-dependent parameter, and $a_j$ is a stellar dependent parameter.
After this first pass, the VLT data has a per-point scatter of 0.13\%,
down from 0.27\%.  Any remaining trends in the target are further reduced
by performing a multiple linear regression fit with position in the
detector, FHWM of the stars and airmass.  This second detrending pass
further reduces the scatter of the VLT light curve to 0.116\%, close
to the Poisson limit of $\sim$0.10\%.  For the Magellan data, the
initial per-point scatter of the light curve was 0.16\% and a final
scatter of 0.098\% was achieved after de-trending, close to the
Poisson limit of $\sim$0.08\%.  The optimal VLT and
Magellan apertures, and the sky background annulus for the Magellan data,
were determined for the target and reference stars which minimized the
standard deviation of the light curve when using the apertures in the above
de-trending procedures. The final light curves are shown in Figure 1.

\begin{figure}
{\centering	 
  \includegraphics[width=0.47\textwidth,angle=0]{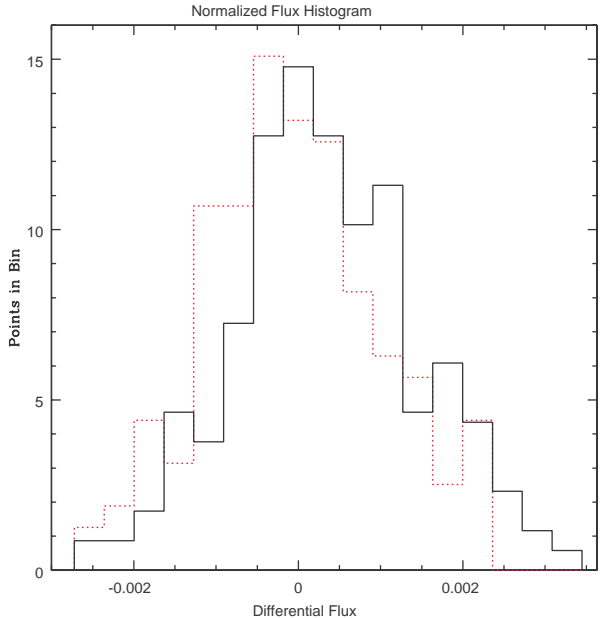}}
\caption[]{ Normalized flux histograms for the portions of the combined VLT and 
Magellan data in-eclipse (dotted line) and out of eclipse (solid line). The bin 
width is 0.000363, coincident with the detected transit depth.}

\end{figure}

\subsection{Eclipse detection and error estimation}

The VLT and Magellan datasets combined contain 596 points between
phases 0.3965 and 0.6403, based on the transit ephemerides equation of
\citet{2008ApJ...677.1324T}. Given the photometric precisions per data
point of 0.116$\%$ for the VLT data and 0.098$\%$ for the Magellan data,
the 0.05$\%$ maximum expected depth of the thermal emission signature
of OGLE-TR-56b during secondary eclipse is at least a factor of two
shallower than the errors of the individual points. Therefore, the
eclipse is difficult to discern by eye in the unbinned data (see Fig.~1) and
further tests are necessary to confirm its presence.

First, we used the known orbital period and stellar and planetary
radii from \citet{2008ApJ...677.1324T} to fit transit models
\citep{2002ApJ...580L.171M} with no limb darkening to the data from
each night.  The best fits were found using a Levenberg-Marquardt
least-square algorithm, with the central phase of the transit set to
0.5, consistent with a circular orbit \citep{2005A&A...431.1105B}, and
leaving the depth and the flux out of transit as free parameters. The
results are a transit depth of 0.037$\pm$0.016$\%$ with a reduced
$\chi^{2}$ of 0.903 for VLT and 0.036$\pm$0.011$\%$ with a reduced
$\chi^{2}$ of 0.926 for Magellan. The total errors in the eclipse
depth are estimated as $\sigma_{depth}^2=\sigma_w^2/N+\sigma_r^2$
where $\sigma_w$ is the scatter per-point out-of-transit and
$\sigma_r^2$ describes the red noise \citep{2006MNRAS.373..231P}, estimated with the
``prayer-bead'' method (see \citealt{2007A&A...471L..51G,
2008MNRAS.386.1644S}), to be 1.1$\times10^{-4}$ in the VLT data and
4$\times10^{-5}$ in the Magellan data.

\begin{figure}
{\centering	 
  \includegraphics[width=0.36\textwidth,angle=90]{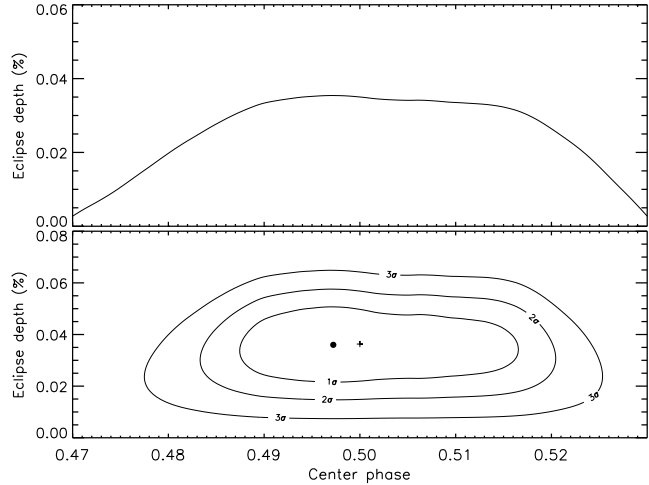}}
\caption[]{The top panel shows the derived model eclipse depth versus
assumed central transit phase, with the plot approximately covering
the eclipse duration.  Phase 0.5 corresponds to a circular orbit.  The
central phase is not well constrained within approximately 0.01, with
similar transit depths found over a phase range of 0.49 to 0.515, but
the transit depth falls off rapidly for central phases outside that
range. The best-fit value (black dot at phase 0.497) is indicated in
the bottom contour plot, along with the value assuming a circular
orbit (cross).  The joint confidence contours of the best fit are also
plotted in the bottom figure for the 68.3\%, 95.5\% and 99.7\%
levels.}
\end{figure}

We finally investigated to what extent uncertainties in the
system's parameters affect our eclipse depth measurements. Varying the
impact parameter, inclination, planet-to-star radius ratio, and
system scale from \citet{2008ApJ...677.1324T} by 1$\sigma$, the
measured eclipse depth in the Magellan data changed by only
$\pm$2.4$\times10^{-5}$ or 0.22 $\sigma_{depth}$.  Thus, our result
is largely independent of the system parameters

Three other tests were performed to confirm the detection. Taking the
average of the 365 light curve data points at out-of-transit phases
versus 174 points in-transit (only points near the bottom of the
transit), we measure a depth of 0.0363 $\pm$ 0.0091$\%$, which
corresponds to a 3-4$\sigma$ detection including red-noise.  We further checked the
detection in a manner similar to the reported detection of HD209458 at
24$\mu$m by \citet{2005Natur.434..740D}.  Figure 2 shows the first of
those tests, i.e. histograms of the distribution of normalized flux
for both the in-transit and out-of-transit phase
intervals. The distribution of in-occultation points is shifted by
0.0363$\%$ of the flux, as expected from the presence of the
occultation. The results of the second test are shown in Figure
3. Letting the central phase and depth of the model transit vary, we
confirm the depth and find the center of the occultation to be at
phase $0.497^{+0.010} _{-0.006}$, consistent with a near-circular
orbit.


\begin{figure}
{\centering	 
  \includegraphics[width=0.43\textwidth,angle=0]{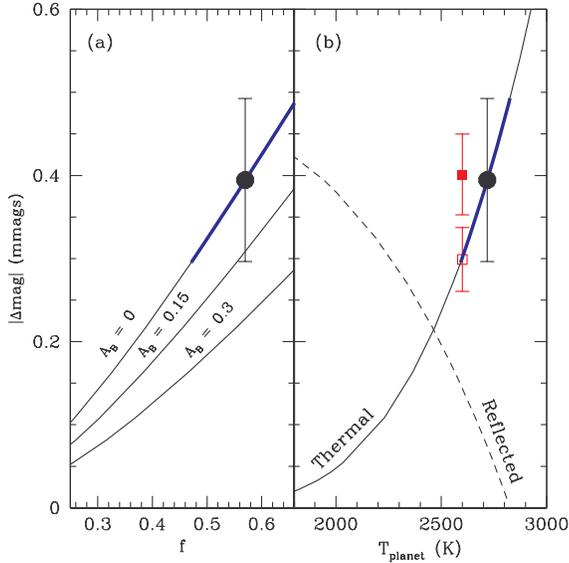}}
\caption[]{ $z'$-band secondary eclipse depth of OGLE-TR-56b as a
function of the reradiation factor $\it f$ for albedos $A_{B}$ = 0, 0.15 and 0.3 (left-side),
and as a function of the planet's temperature (right-side), in the
case of reflected light (dashed line), or from thermal emission
assuming the planet emits as a black-body (solid line). The (red)
filled and open squares indicate the expected model depth of the
transit with and without TiO/VO \citep{2003ApJ...594.1011H}. Our
measured depth is the black dot.}
\end{figure}


\section{Discussion}

A decrease in flux of 0.0363$\pm$0.0091$\%$ corresponds to a
contribution from the planet to the total z'-band brightness of the
system of 0.394$\pm$0.098 milli-magnitudes.  The light detected could
be stellar light reflected by the surface of the planet, but planetary
thermal emission is the best explanation. Theoretical work on Very-Hot
Jupiter atmospheres
\citep{1999ApJ...513..879M,2000ApJ...540..504S,2000ApJ...538..885S}
predict that these planets can reflect at most 30$\%$ of the stellar
incident light (i.e. $A_{B}$ = 0.3), if their atmospheres are covered
by homogeneous pure silicate clouds at altitudes of one millibar or
higher. $A_{B}$ will be less than 0.3 if the planets are cloud free or
have patchy silicate or iron clouds so stellar radiation penetrates
through the clouds and is absorbed by gaseous molecules
\citep{2008MNRAS.389..257H}.  Reflected light could make OGLE-TR-56b
appear bright enough to be detected in the z'-band, but this requires
$A_{B}$ $>>$ 0.3, and Very-Hot Jupiter atmospheres are too hot on the
substellar side for silicate or iron clouds to form, implying low
albedos.  The measured radiation most likely comes from thermal
emission and the effective temperature of the planet has to be high.
We estimate a brightness temperature of $T_{z'}$ =
$2718^{+127}_{-107}$ K, based on the eclipse depth error bars, and
assuming the planet emits thermally as a black-body (see Figure
4b). $A_{B}=0$ and $f > 0.47$ are also values consistent with the eclipse depth (See Figure 4a). This indicates that the stellar
radiation is absorbed by the planet and not efficiently distributed
throughout its atmosphere via winds, but instead is almost instantaneously
re-radiated back to space \citep{2007ApJ...667L.191L}.  The blackbody $T_{z'}$
of the planet would be 2260$\pm$19 K if the incident stellar radiation 
were completely redistributed and $A_{B}=0$.


Recent detections of planetary occultations at infrared wavelengths
($\sim$ 3--24 $\mu$m) by the Spitzer space telescope have revealed
properties of the atmospheric structure of hot-Jupiter exoplanets,
most notably thermal inversions \citep{2003ApJ...594.1011H,
2007ApJ...668L.171B,2007Natur.447..691H,
2008ApJ...673..526K,2008arXiv0808.1913S}, thought to be caused by a
strong optical absorber at high altitudes
\citep{2007ApJ...668L.171B,2008ApJ...678.1419F}. Hot Jupiters with
thermal inversions have bright near-infrared and infrared emission
lines where the planetary flux is much greater than a single
temperature black-body model would predict.  OGLE-TR-56b is hotter than
any of the planets Spitzer has observed, and at the extreme hot and
highly-irradiated end of those planets believed to have inversions.
While our result in Fig.~4 reveals a depth closest to the expected value from
the \citet{2003ApJ...594.1011H} model with thermal inversion, more
observations of the secondary transit of OGLE-TR-56b are necessary to
improve our current error bars and confirm this result. 
Also, the contrast between models with and without thermal inversion
is greater in the near-infrared, with predicted occultation depths of
1--3 mmags \citep{2008ApJ...678.1419F}, making detections possible
with modern ground-based detectors. Detecting the secondary transit of
this planet in the near-infrared will also further constrain the
models.

 All secondary transit measurements to date have been made with
Spitzer.  A ground-based detection now extends exoplanet occultation
science to shorter wavelengths, more accessible telescopes, and
fainter targets.  This is particularly important for the near
future, as Spitzer is soon to run out of cryogens, severely limiting
its capabilities.  The eventual loss of Spitzer will leave a long time
window before the launch of the James Webb Space Telescope, when
occultation measurements will rely heavily on ground-based
measurements.  At the same time, planet-finding missions like CoRoT
and Kepler will find a wealth of new planets to study.  Extending
ground-based measurements to other very hot Jupiters will enable
comparative exoplanetology, where the atmospheric structures of hot
planets can be compared and contrasted, to continue 
investigation into important and relevant physical processes.

\begin{acknowledgements}
D.K.S. is supported by CNES. MLM acknowledges support provided by NASA
through Hubble Fellowship grant HF-01210.01-A awarded by the STScI,
which is operated by the AURA, Inc. for NASA, under contract
NAS5-26555.  We wish to thank J. Elliot, P.I. of MagIC-E2V, for
providing access to the instrument, S. Seager, F. Pont and A. Lecavelier for helpful
comments, suggestions and discussions, and the Magellan staff and ESO
staff at Paranal.

\end{acknowledgements}


\bibliographystyle{aa} 

\begin{thebibliography}{24}
\expandafter\ifx\csname natexlab\endcsname\relax\def\natexlab#1{#1}\fi

\bibitem[{{Bouchy} {et~al.}(2005){Bouchy}, {Pont}, {Melo}, {Santos}, {Mayor},
  {Queloz}, \& {Udry}}]{2005A&A...431.1105B}
{Bouchy}, F., {Pont}, F., {Melo}, C., {et~al.} 2005, \aap, 431, 1105

\bibitem[{{Burrows} {et~al.}(2007){Burrows}, {Hubeny}, {Budaj}, {Knutson}, \&
  {Charbonneau}}]{2007ApJ...668L.171B}
{Burrows}, A., {Hubeny}, I., {Budaj}, J., {Knutson}, H.~A., \& {Charbonneau},
  D. 2007, \apjl, 668, L171

\bibitem[{{Deming} {et~al.}(2007){Deming}, {Richardson}, \&
  {Harrington}}]{2007MNRAS.378..148D}
{Deming}, D., {Richardson}, L.~J., \& {Harrington}, J. 2007, \mnras, 378, 148

\bibitem[{{Deming} {et~al.}(2005){Deming}, {Seager}, {Richardson}, \&
  {Harrington}}]{2005Natur.434..740D}
{Deming}, D., {Seager}, S., {Richardson}, L.~J., \& {Harrington}, J. 2005,
  \nat, 434, 740

\bibitem[{{Fortney} {et~al.}(2008){Fortney}, {Lodders}, {Marley}, \&
  {Freedman}}]{2008ApJ...678.1419F}
{Fortney}, J.~J., {Lodders}, K., {Marley}, M.~S., \& {Freedman}, R.~S. 2008,
  \apj, 678, 1419

\bibitem[{{Gillon} {et~al.}(2007){Gillon}, {Demory}, {Barman}, {Bonfils},
  {Mazeh}, {Pont}, {Udry}, {Mayor}, \& {Queloz}}]{2007A&A...471L..51G}
{Gillon}, M., {Demory}, B.-O., {Barman}, T., {et~al.} 2007, \aap, 471, L51

\bibitem[{{Harrington} {et~al.}(2007){Harrington}, {Luszcz}, {Seager},
  {Deming}, \& {Richardson}}]{2007Natur.447..691H}
{Harrington}, J., {Luszcz}, S., {Seager}, S., {Deming}, D., \& {Richardson},
  L.~J. 2007, \nat, 447, 691

\bibitem[{{Hood} {et~al.}(2008){Hood}, {Wood}, {Seager}, \& {Collier
  Cameron}}]{2008MNRAS.389..257H}
{Hood}, B., {Wood}, K., {Seager}, S., \& {Collier Cameron}, A. 2008, \mnras,
  389, 257

\bibitem[{{Hubeny} {et~al.}(2003){Hubeny}, {Burrows}, \&
  {Sudarsky}}]{2003ApJ...594.1011H}
{Hubeny}, I., {Burrows}, A., \& {Sudarsky}, D. 2003, \apj, 594, 1011

\bibitem[{{Knutson} {et~al.}(2008){Knutson}, {Charbonneau}, {Allen}, {Burrows},
  \& {Megeath}}]{2008ApJ...673..526K}
{Knutson}, H.~A., {Charbonneau}, D., {Allen}, L.~E., {Burrows}, A., \&
  {Megeath}, S.~T. 2008, \apj, 673, 526

\bibitem[{{Knutson} {et~al.}(2007){Knutson}, {Charbonneau}, {Deming}, \&
  {Richardson}}]{2007PASP..119..616K}
{Knutson}, H.~A., {Charbonneau}, D., {Deming}, D., \& {Richardson}, L.~J. 2007,
  \pasp, 119, 616

\bibitem[{{L{\'o}pez-Morales} \& {Seager}(2007)}]{2007ApJ...667L.191L}
{L{\'o}pez-Morales}, M. \& {Seager}, S. 2007, \apj, 667, L191

\bibitem[{{Mandel} \& {Agol}(2002)}]{2002ApJ...580L.171M}
{Mandel}, K. \& {Agol}, E. 2002, \apj, 580, L171

\bibitem[{{Marley} {et~al.}(1999){Marley}, {Gelino}, {Stephens}, {Lunine}, \&
  {Freedman}}]{1999ApJ...513..879M}
{Marley}, M.~S., {Gelino}, C., {Stephens}, D., {Lunine}, J.~I., \& {Freedman},
  R. 1999, \apj, 513, 879

\bibitem[{{Pont} {et~al.}(2006){Pont}, {Zucker}, \&
  {Queloz}}]{2006MNRAS.373..231P}
{Pont}, F., {Zucker}, S., \& {Queloz}, D. 2006, \mnras, 373, 231

\bibitem[{{Redfield} {et~al.}(2008){Redfield}, {Endl}, {Cochran}, \&
  {Koesterke}}]{2008ApJ...673L..87R}
{Redfield}, S., {Endl}, M., {Cochran}, W.~D., \& {Koesterke}, L. 2008, \apjl,
  673, L87

\bibitem[{{Seager} {et~al.}(2008){Seager}, {Deming}, \&
  {Valenti}}]{2008arXiv0808.1913S}
{Seager}, S., {Deming}, D., \& {Valenti}, J.~A. 2008, ArXiv.0808.1913

\bibitem[{{Seager} {et~al.}(2000){Seager}, {Whitney}, \&
  {Sasselov}}]{2000ApJ...540..504S}
{Seager}, S., {Whitney}, B.~A., \& {Sasselov}, D.~D. 2000, \apj, 540, 504

\bibitem[{{Snellen}(2005)}]{2005MNRAS.363..211S}
{Snellen}, I.~A.~G. 2005, \mnras, 363, 211

\bibitem[{{Snellen} \& {Covino}(2007)}]{2007MNRAS.375..307S}
{Snellen}, I.~A.~G. \& {Covino}, E. 2007, \mnras, 375, 307

\bibitem[{{Southworth}(2008)}]{2008MNRAS.386.1644S}
{Southworth}, J. 2008, \mnras, 386, 1644

\bibitem[{{Sudarsky} {et~al.}(2000){Sudarsky}, {Burrows}, \&
  {Pinto}}]{2000ApJ...538..885S}
{Sudarsky}, D., {Burrows}, A., \& {Pinto}, P. 2000, \apj, 538, 885

\bibitem[{{Tamuz} {et~al.}(2005){Tamuz}, {Mazeh}, \&
  {Zucker}}]{2005MNRAS.356.1466T}
{Tamuz}, O., {Mazeh}, T., \& {Zucker}, S. 2005, \mnras, 356, 1466

\bibitem[{{Torres} {et~al.}(2008){Torres}, {Winn}, \&
  {Holman}}]{2008ApJ...677.1324T}
{Torres}, G., {Winn}, J.~N., \& {Holman}, M.~J. 2008, \apj, 677, 1324

\end{thebibliography}




\end{document}